\documentclass[11pt]{article}

\usepackage{fullpage,parskip}
\usepackage[utf8]{inputenc} % allow utf-8 input
\usepackage[T1]{fontenc}    % use 8-bit T1 fonts
\usepackage[hidelinks,colorlinks=true,citecolor=black!30!blue,linkcolor=black!10!red]{hyperref}       % hyperlinks
\usepackage{url}            % simple URL typesetting
\usepackage{enumitem}
\usepackage{booktabs}       % professional-quality tables
\usepackage{amsfonts}       % blackboard math symbols
\usepackage{nicefrac}       % compact symbols for 1/2, etc.
\usepackage{microtype}      % microtypography
\usepackage{xcolor}         % colors

\usepackage{hyperref}
\usepackage{graphicx}

\usepackage{wrapfig}
\usepackage[utf8]{inputenc} % allow utf-8 input
\usepackage[T1]{fontenc}    % use 8-bit T1 fonts
\usepackage{hyperref,xcolor}       % hyperlinks
\usepackage{url}            % simple URL typesetting
\usepackage{booktabs}       % professional-quality tables
\usepackage{amsfonts}       % blackboard math symbols
\usepackage{nicefrac}       % compact symbols for 1/2, etc.
\usepackage{microtype}      % microtypography
\usepackage{xcolor}         % colors
\usepackage{comment}

\usepackage{amsmath,amsfonts,amssymb}
\usepackage{graphicx}

\usepackage{algorithm}
\usepackage{algorithmic}

\usepackage{multirow}

\usepackage{bm}
\usepackage[all]{hypcap}
\usepackage{caption}
\usepackage{subcaption}
\usepackage{mathtools} % for xlongleftrightarrow
\usepackage{mleftright}  % fixes some annoying spacing issues

\usepackage{amsthm, thmtools}
\usepackage{amsmath, amsfonts, mathtools, amssymb}
\usepackage{float}
\newtheoremstyle{sltheorem}
                {}          % Space above
                {}          % Space below
                {\slshape}  % Body font
                {}          % Indent amount
                {\bfseries} % Head font
                {.}         % Punctuation after head
                { }         % Space after theorem head
                {}          % Theorem head spec
\theoremstyle{sltheorem}

\DeclareMathOperator*{\argmin}{arg\,min}

\usepackage{etoolbox}
\usepackage[toc,page]{appendix}

\usepackage{xspace}
\usepackage{bbm}

\def\tr{\mathop{\rm tr}\nolimits}%
\def\diag{\mathop{\rm diag}\nolimits}%
\def\Re{\mathop{\rm Re}\nolimits}%

\newcommand{\Xb}{\mathbf{X}}

\newcommand{\wv}{{\bf w}}

\newcommand{\Ac}{\mathcal{A}}

\newcommand{\Nc}{\mathcal{N}}

\newcommand{\Xc}{\mathcal{X}}

\newcommand{\gv}{\bm{g}}

\newcommand{\av}{{\bf a}}
\newcommand{\xv}{{\bf x}}
\newcommand{\yv}{{\bf y}}
\newcommand{\zv}{{\bf z}}
\newcommand{\uv}{{\bf u}}
\newcommand{\vv}{{\bf v}}
\newcommand{\rv}{{\bf r}}
\newcommand{\sv}{{\bf s}}

\newcommand{\hv}{{\bf h}}
\newcommand{\qv}{{\bf q}}

\DeclareMathOperator\E{E}

\def\textiid{i.i.d.\@\xspace}
\newcommand\iid{\ifmmode\text{ i.i.d. } \else \textiid \fi}

\title{Maximum Likelihood Reconstruction for Multi-Look Digital Holography with Markov-Modeled Speckle Correlation}

\author{Xi Chen, Arian Maleki, Shirin Jalali}

\date{}

\begin{document}

\maketitle

\renewcommand\thefootnote{}
\footnotetext{
Xi Chen and Shirin Jalali are with the Department of Electrical and Computer Engineering,
Rutgers University, Piscataway, NJ 08854 USA
(e-mail: xi.chen15@rutgers.edu; shirin.jalali@rutgers.edu).}

\footnotetext{
Arian Maleki is with the Department of Statistics,
Columbia University, New York, NY 10027 USA
(e-mail: arian@stat.columbia.edu).
}

\begin{abstract}
Multi-look acquisition is a widely used strategy for reducing speckle noise in coherent imaging systems such as digital holography. By acquiring multiple measurements, speckle can be suppressed through averaging or joint reconstruction, typically under the assumption that speckle realizations across looks are statistically independent. In practice, however, hardware constraints limit measurement diversity, leading to inter-look correlation that degrades the performance of conventional methods. In this work, we study the reconstruction of speckle-free reflectivity from complex-valued multi-look measurements in the presence of correlated speckle. We model the inter-look dependence using a first-order Markov process and derive the corresponding likelihood under a first-order Markov approximation, resulting in a constrained maximum likelihood estimation problem. To solve this problem, we develop an efficient projected gradient descent framework that combines gradient-based updates with implicit regularization via deep image priors, and leverages Monte Carlo approximation and matrix-free operators for scalable computation. Simulation results demonstrate that the proposed approach remains robust under strong inter-look correlation, achieving performance close to the ideal independent-look scenario and consistently outperforming methods that ignore such dependencies. These results highlight the importance of explicitly modeling inter-look correlation and provide a practical framework for multi-look holographic reconstruction under realistic acquisition conditions. Our code is available at: \url{https://github.com/Computational-Imaging-RU/MLE-Holography-Markov}.
\end{abstract}

\section{Introduction}\label{sec:introduction}

Speckle noise is one of the primary challenges in coherent imaging systems, such as holographic displays \cite{yoo2021optimization, schiffers2023stochastic, schiffers2025holochrome} and wavefront imaging through scattering media \cite{wu2019wish}, as well as digital holography \cite{bianco2018strategies,pellizzari2017phase}, severely degrading the perceived image quality. In such systems, speckle arises from the interference of coherent waves undergoing random phase variations, which can be induced by both surface roughness and propagation through scattering media. These effects lead to granular intensity fluctuations in the recorded image. Over the decades, several optical and computational approaches for mitigating the impact of speckle noise have been developed \cite{bianco2018strategies}. One of the most effective methods is the so-called multi-look acquisition, which acquires multiple sets of measurements and generates the reconstructed image by properly averaging or jointly processing these measurements. If the speckle realizations are independent across the looks, this method leads to a reduction in the impact of speckle noise, and an improvement in reconstructed image quality.

Multi-look measurement is typically achieved by introducing controlled diversity in the optical field, such that each measurement corresponds to an independent (or weakly correlated) speckle pattern while preserving the same reflectivity information. For instance, in holographic displays, temporal multiplexing~\cite{choi2022time} generates a sequence of phase patterns over time, while polarization multiplexing~\cite{nam2023depolarized} exploits orthogonal polarization states to produce statistically independent speckle realizations. These approaches have been empirically validated as effective in reducing speckle noise \cite{amako1995speckle, rong2010speckle, yoo2021optimization, schiffers2023stochastic, lee2024speckle, schiffers2025holochrome} in holographic systems.

While ideally one requires independent speckle noise realizations across looks, in practice, achieving full independence remains challenging \cite{bianco2018strategies}. The diversity introduced by spatial light modulators, phase masks, or diffusers can be fundamentally limited by finite spatial resolution, phase quantization, and hardware imperfections \cite{yang2023review}. As a result, the speckle realizations across different looks are not always guaranteed to be independent, which limits the performance gains promised by the multi-look strategy. 
Modeling of dynamic speckle has been studied from multiple perspectives in the literature. For instance, in \cite{burrell2021wave1,burrell2021wave2}, the authors consider deterministic geometric motion models and analytically characterize how such motion affects speckle correlation, enabling explicit decorrelation of dynamic speckle. In a different line of work, the temporal statistics of speckle are treated as informative signals, where temporal correlation is estimated to quantify motion dynamics, such as flow or diffusion \cite{bandyopadhyay2005speckle,zilpelwar2022model}.

In this work, we propose a stochastic model that captures the dependencies between speckle realizations across looks. Based on this model, we formulate a constrained maximum likelihood estimation (MLE) problem, which results in a computationally challenging non-convex optimization. Focusing on digital holography, we develop an efficient Monte Carlo–based algorithm to solve the resulting MLE problem. Simulations demonstrate that the proposed approach improves reconstruction performance over an MLE baseline that ignores inter-look dependencies under correlated measurements, while maintaining scalability to high-resolution imaging systems through efficient, matrix-free implementations.

% \textcolor{blue}{positive parts of multi-look, it is effective, but the prerequsite is the access to the independent looks, which is not always guaranteed in practice. Even though, can we statistically model the correlation, and algorithmically solve it.} 

% \textcolor{blue}{move this to the correlation part}In practice, such decorrelation can be induced through mechanisms including rotating diffusers, phase modulation, wavelength or angle diversity, and polarization changes. 

\section{Digital holography: Problem setting}
We begin by reviewing the mathematical model of the multi-look digital holography system. We then formulate the corresponding MLE problem and discuss its algorithmic solutions, along with the associated challenges and limitations of existing approaches. Based on this analysis, we conclude by summarizing the contributions of this work.
\subsection{Mathematical Model}
% The early works including \cite{sotthivirat2004penalized, lee2016single} focus on reconstructing a complex-valued object field directly from real-valued hologram measurements. These approaches primarily address phase retrieval and wavefront reconstruction. In contrast, off-axis digital holography provides access to the complex-valued wavefront, from which the recent works \cite{pellizzari2017phase,pellizzari2019imaging,pellizzari2020coherent,bate2021model,allen2025clamp} seek to estimate the underlying speckle-free reflectivity by modeling the object field as a random process with spatially varying variance, thereby explicitly accounting for speckle noise, as below 

A multi-look coherent imaging system is mathematically described as follows: for $\ell=1,\ldots,L$, 
\begin{align} \label{eq:forward_multi}
    \yv_{\ell} = A \mathbf{g}_{\ell} + \zv_{\ell},
\end{align}
where $\yv_{\ell} \in \mathbb{C}^m$ is the measurement vector, $A \in \mathbb{C}^{n \times n}$ denotes the  sensing matrix, and $\zv_{\ell} \sim \mathcal{CN}(\mathbf{0},\sigma_z^2 I_m)$ is the additive sensor noise. Let $X = \diag(\xv)$ denote the diagonal matrix formed from the underlying speckle-free reflectivity $\xv$. Given $\xv$, ideally, for $\ell=1, \ldots, L$, the vectors $\mathbf{g}_{\ell} \sim \mathcal{CN}(\mathbf{0}, X) \in \mathbb{C}^{n}$ are independent and represent the complex field after random phase interference, where the randomness models fully developed speckle arising from coherent interference \cite{zhou2022compressed,chenbagged}.

The model described above applies to general coherent imaging systems. In this paper, however, we focus on digital holography, which corresponds to a specific form of the forward model and will be important for our subsequent algorithmic developments. In digital holography, based on the Fresnel model, the sensing matrix can be written as
\begin{align} \label{eq:sensing_matrix}
A = \mathcal{F}^{-1} M \mathcal{F},
\end{align}
where $\mathcal{F} \in \mathbb{C}^{n \times n}$ denotes the discrete Fourier transform operator. By Fourier optics \cite{goodman2005introduction}, the Fourier transform is applied to the 2D image. When acting on the vectorized image $\xv$, the operator $\mathcal{F}$ can be written as the Kronecker product of row and column Fourier matrices, i.e.,
\[
\mathcal{F} = \mathcal{F}_r \otimes \mathcal{F}_c^H.
\]
The matrix $M = \mathrm{diag}(\mathrm{vec}(P)) \in \{0,1\}^{n \times n}$ represents the aperture mask in the spatial frequency domain, where $P \in \{0,1\}^{H \times W}$ denotes the 2D aperture function.

The complex optical field $\yv_{\ell}$ is obtained either directly via off-axis digital holography or through wavefront reconstruction using phase retrieval algorithms \cite{latychevskaia2019iterative}. In off-axis digital holography, the interference between the object and a tilted reference beam shifts the signal of interest away from the zero-frequency component in the Fourier domain. This enables the separation of the desired complex field through appropriate demodulation and filtering operations, typically by isolating the first-order diffraction term via bandpass filtering and then applying an inverse Fourier transform to recover the complex-valued field. We then formulate a constrained MLE problem of reconstructing $\xv$ from the measurements $\yv$ with known sensing matrix $A$.

\subsection{Maximum Likelihood Estimation}\label{sec:2-2}

Given the complex-valued measurements $\yv_1,\ldots,\yv_{L}$ and the sensing matrix $A$, the goal of the recovery algorithm is to estimate the underlying reflectivity $\xv$. Note that although the forward operator $A$ in digital holography is typically square, the presence of an aperture mask with limited transparency generally renders it rank-deficient, thereby making the resulting inverse problem ill-posed \cite{pellizzari2020coherent, chen2026monte}.

Given the independence and Gaussianity of  $\zv_{\ell}$ and $\gv_{\ell}$, with fixed $\xv$ and $A$, the measurement vectors $\yv_1,\ldots,\yv_L$ are independently distributed as zero-mean Gaussian random vectors with  covariance  $\Sigma(\xv)$ defined as 
\begin{gather}
     \Sigma(\xv)= AXA^H + \sigma_z^2 I_m.
\end{gather}
Given $\xv$, the  negative log-likelihood of  $\yv_\ell$, $ -\log p(\yv_\ell|A,X)$, denoted as $f_\ell(\xv)$, up to  additive constants independent of $\xv$,  can be written as
\[
f_\ell(\xv)  = \log \det \Sigma(\xv) + \yv_\ell^H \Sigma^{-1}(\xv) \yv_\ell,
\]
where $\yv_\ell^H$ is the Hermitian transpose of $\yv_\ell$. Given the assumption on the independence of the measurements, the normalized negative log-likelihood of all  $L$ looks can be   written as $f(\xv)={\frac{1}{L}}\sum_{\ell}f_{\ell}(\xv)$. That is,
\begin{align} \label{eq:objective_independent}
f(\xv) = \log \det \Sigma(\xv) + \frac{1}{L}\sum^L_{\ell=1} \yv_\ell^H \Sigma^{-1}(\xv) \yv_\ell.
\end{align}
It is known that properly modeling and exploiting the underlying source structure plays a key role in solving the noisy and ill-posed inverse problems, and can substantially improve reconstruction performance. So, instead of optimizing over all $\xv \in (\mathbb{R}^+)^n$, we assume that $\xv$ belongs to a class of images $\Xc$ that captures prior knowledge of the desired signal. This leads to the constrained MLE formulation
\begin{align}
    \hat{\xv} = \argmin_{\xv \in \Xc} f(\xv).
\end{align}
The choice of $\Xc$ plays a central role in regularizing the reconstruction and mitigating the effects of speckle noise and limited measurements.

% However, the statistical nature of speckle is rarely explicitly modeled in the reconstruction algorithms, especially in conjunction with physically accurate forward models. 

Prior work \cite{pellizzari2017phase,pellizzari2019imaging,pellizzari2020coherent,bate2021model} models speckle statistically as in \eqref{eq:forward_multi}, and recovers the reflectivity $\xv$ from the complex field $\yv$ via maximum likelihood estimation, often within a Plug-and-Play (PnP) framework \cite{venkatakrishnan2013plug} to incorporate learned image priors such as DnCNN \cite{zhang2017beyond}. Due to the intractability of directly optimizing the likelihood, these approaches rely on variational approximations and EM-based optimization \cite{dempster1977maximum}. In addition, simplifying assumptions on the sensing matrix (e.g., $A^H A = I$) are often introduced. While effective in practice, these approximations can lead to suboptimal reconstruction performance.

More recently, \cite{zhou2022compressed,chenbagged} theoretically characterize the performance of MLE and show that accurate recovery is possible despite the ill-posedness of the problem. To address the computational burden of matrix inversion in the MLE objective, \cite{chenbagged} employs the Newton-Schulz iterative algorithm \cite{gower2017randomized,stotsky2020efficient}, enabling direct optimization of the likelihood. For digital holography, \cite{chen2025monte, chen2026monte} propose a Monte Carlo-based approach for matrix diagonal approximation, combined with conjugate gradient methods, which significantly accelerates gradient computation and enables scalable, high-resolution reconstruction with exact likelihood modeling and accurate aperture representation.

These works typically assume that the acquired multi-look measurements correspond to independent speckle noise realizations. In practice, however, existing decorrelation methods are imperfect, and non-negligible correlation across looks often persists. This raises a fundamental question: can we explicitly model the statistical dependence across multiple measurements and leverage it to develop an improved MLE-based algorithm?

%\textcolor{blue}{Meanwhile, the assumption on the independence across the looks simplifies the MLE-based optimization \cite{bate2021model,chen2026monte}. Given the inter-look correlation, a gap is left between the existing optimization algorithm and the multi-look measurements model which depicts the correlation. }

\subsection{Contributions}

In this paper we focus on addressing the question raised at the end of Section~\ref{sec:2-2}. The main contributions of this work are as follows:
\begin{itemize}
    \item We model the dependence across speckle realizations in multi-look digital holography using a first-order Markov process, and incorporate this structure into the likelihood formulation alongside the Fresnel imaging model.
    
    \item We formulate the corresponding constrained maximum likelihood estimation problem that explicitly accounts for inter-look correlation.
    
    \item We develop an efficient algorithm to solve the resulting optimization, including the estimation of the correlation coefficients across looks, while preserving accurate aperture modeling and enabling scalability to high-resolution settings.
\end{itemize}
We then demonstrate the effectiveness of the proposed method through simulation results.

\section{Method}

We begin by developing a stochastic model for the inter-look speckle correlation in multi-look coherent imaging. We then derive the corresponding likelihood function based on a first-order approximation of this correlation structure. Focusing on digital holography, we address the computational challenges introduced by the resulting objective function by proposing an efficient algorithm that combines Monte Carlo approximation with the conjugate gradient method.

\subsection{Markov Model of Speckle Correlation in Multi-Look}

As discussed earlier, in practical acquisition setups, speckle realizations across multiple looks are not strictly independent. To mathematically characterize this inter-look correlation and study its impact on reconstruction performance, we model the speckle field across looks as a first-order Markov process.

Consider a sequence of $L$ looks as defined in \eqref{eq:forward_multi}. We assume that the speckle corresponding to the first look is fully developed \cite{goodman2007speckle}, i.e., $\gv_1 \sim \mathcal{CN}(\mathbf{0}, X)$. For $\ell = 2, \ldots, L$, to capture the dependence between successive speckle realizations, we let $\gv_\ell$ evolve according to
\begin{align} \label{eq:speckle_corre}
\gv_\ell = \alpha \gv_{\ell-1} + \sqrt{1-\alpha^2}\,\uv_\ell,
\end{align}
where, for a fixed $\xv$, the random vectors $\uv_1, \ldots, \uv_L$ are independent of $\gv_1$ and are independently and identically distributed as $\uv_\ell \sim \mathcal{CN}(\mathbf{0}, X)$. Here, $\alpha \in [0,1]$ denotes the correlation coefficient between consecutive looks. The case $\alpha = 0$ corresponds to fully independent speckle realizations. This construction ensures that $\gv_\ell \sim \mathcal{CN}({\bf 0}, X)$, $\mathbb{E}[\gv_\ell] = \xv$ for all $\ell$, while introducing controlled correlation across looks. Note that, for $\ell=2,\ldots,L$,
\begin{align}
\gv_\ell = \alpha^{\ell-1} \gv_{1} + \sqrt{1 - \alpha^{2(\ell-1)}}\, \tilde{\uv},
\end{align}
where $\tilde{\uv} \sim \mathcal{CN}({\bf 0}, X)$ is independent of $\gv_1$.

\subsection{Likelihood of Correlated Speckle}

Recall that, given the speckle correlation in \eqref{eq:speckle_corre}, the multi-look measurement process considering the sensing matrix $A$ is given by $\yv_1 = A \gv_1 + \zv_1$, and, for $\ell = 2, \ldots, L$,
\begin{align}
\yv_\ell = A \gv_\ell + \zv_\ell = \alpha A \gv_{\ell-1} + \sqrt{1-\alpha^2}\, A \uv_\ell + \zv_\ell,
\label{eq:forward_model_correlation}
\end{align}
where $\uv_2,\ldots,\uv_L \stackrel{\rm i.i.d.}{\sim} \mathcal{CN}({\bf 0}, X)$ and $\zv_1,\ldots,\zv_L \stackrel{\rm i.i.d.}{\sim} \mathcal{CN}({\bf 0}, \sigma_z^2 I)$.

Unlike the case of independent speckle, the measurements $\yv_1,\ldots,\yv_L$ are statistically dependent given $\xv$. Moreover, due to the presence of additive noise, the sequence $\{\yv_\ell\}_{\ell=1}^L$ does not form a first-order Markov process. As a result, directly characterizing the full likelihood function is intractable.

To address this challenge, we instead adopt a first-order approximation of the joint distribution. Specifically, rather than expressing the likelihood as
\begin{align}
    p(\yv_1,\ldots,\yv_L\mid A, X)  = p(\yv_1\mid A, X)\prod_{\ell=2}^L p(\yv_\ell \mid \yv_{\ell-1}, \ldots, \yv_1, A, X),
\end{align}
we approximate it using a first-order Markov model as
\begin{align}
    p(\yv_1\mid A, X)\prod_{\ell=2}^L p(\yv_\ell \mid \yv_{\ell-1}, A, X).
\end{align}
Note that in the absence of additive noise, this approximation becomes exact.

 For $\ell \geq 2$,  $(\yv_{\ell-1},\yv_\ell)$ follows a joint complex Gaussian distribution given by \begin{equation} \label{eq:multi_Gaussian}
\begin{bmatrix}
\yv_{\ell-1} \\
\yv_\ell
\end{bmatrix}
\sim
\mathcal{CN} \left(
\begin{bmatrix}
\boldsymbol\mu_{\ell-1} \\
\boldsymbol\mu_\ell
\end{bmatrix},
\begin{bmatrix}
\boldsymbol\Sigma_{(\ell-1)(\ell-1)} & \boldsymbol\Sigma_{(\ell-1)\ell} \\
\boldsymbol\Sigma_{\ell(\ell-1)} & \boldsymbol\Sigma_{\ell \ell}
\end{bmatrix}
\right),
\end{equation}
where 
\begin{align}
\boldsymbol\mu_{\ell} = \boldsymbol\mu_{\ell-1} = \mathbf{0},
\end{align}
and 
\begin{align}
\boldsymbol\Sigma_{(\ell-1)(\ell-1)} &= \boldsymbol\Sigma_{\ell\ell} = B + \sigma_z^2 I_n,\nonumber \\
\boldsymbol\Sigma_{(\ell-1)\ell} &= \boldsymbol\Sigma_{\ell(\ell-1)} = \alpha B,
\end{align}
with
\begin{equation}
B \triangleq A X A^H = \sum_{i=1}^n x_i \av_i \av_i^H.\label{eq:def-B}
\end{equation}
Therefore, the conditional distribution of $\yv_{\ell}$ given $\yv_{\ell-1}$ is also a complex
Gaussian. That is, 
\begin{equation*}
\yv_\ell \mid \yv_{\ell-1}
\sim
\mathcal{CN}\!\left(
\boldsymbol\mu_{\ell|\ell-1},
\boldsymbol\Sigma_{\ell|\ell-1}
\right),
\end{equation*}
where the conditional mean and covariance can be characterized as 
\begin{align}
&\boldsymbol\mu_{\ell|\ell-1}=
\boldsymbol\mu_\ell
+
\boldsymbol\Sigma_{\ell(\ell-1)}\boldsymbol\Sigma_{(\ell-1)(\ell-1)}^{-1}
(\yv_{\ell-1} - \boldsymbol\mu_{\ell-1}) = \alpha B (B + \sigma^2_z I_n)^{-1} \yv_{\ell-1},
\\
&\boldsymbol\Sigma_{\ell|\ell-1}
=
\boldsymbol\Sigma_{\ell\ell}
-
\boldsymbol\Sigma_{\ell(\ell-1)}
\boldsymbol\Sigma_{(\ell-1)(\ell-1)}^{-1}
\boldsymbol\Sigma_{(\ell-1)\ell} = (B + \sigma^2_z I_n) - \alpha^2 B(B + \sigma^2_z I_n)^{-1}B.
\end{align}

Ignoring the constants with respect to $\xv$, the resulting approximate negative log-likelihood is
\begin{align}
    \log \det (B + \sigma^2_z I_n) + \yv_1^H (B + \sigma^2_z I_n)^{-1} \yv_1 + \sum^L_{\ell=2} \Big[ \log \det \boldsymbol\Sigma_{\ell|\ell-1} 
    + (\yv_\ell - \boldsymbol\mu_{\ell|\ell-1})^{H} \boldsymbol \Sigma_{\ell|\ell-1}^{-1} (\yv_\ell - \boldsymbol\mu_{\ell|\ell-1}) \Big].
\end{align}
We define the resulting objective function for estimating $\xv$ as
\begin{align} 
f_L(\xv)=& \log \det S 
+ \yv_1^H S^{-1} \yv_1 + \sum_{\ell=2}^L
\Big[
\log \det M
+ \rv_\ell^H M^{-1} \rv_\ell
\Big],\label{eq:loss}
\end{align}
where $B$ is defined in \eqref{eq:def-B}, and 
\begin{align}
S &\triangleq B + \sigma_z^2 I_n, \nonumber\\
M &\triangleq S - \alpha^2 B S^{-1} B, \nonumber\\
\rv_\ell &\triangleq \yv_\ell - \alpha B S^{-1} \yv_{\ell-1}.\label{eq:defs-S-M-rvt}
\end{align}

We next develop an efficient algorithm to minimize the resulting non-convex objective, with a focus on scalable gradient computation and effective incorporation of source structure.

\subsection{Constrained MLE-based Reconstruction}
We aim to solve the following constrained MLE problem, while assuming that $\xv$ belongs to a class of images $\Xc$ that captures prior knowledge of the desired signal,
\begin{align}
\label{eq:MLE_constrained}
    \hat{\xv} = \argmin_{\xv \in \Xc} f_L(\xv),
\end{align}
where $f_L$ is defined in \eqref{eq:loss}.
To solve the nonconvex constrained MLE problem in \eqref{eq:MLE_constrained}, we adopt the projected gradient descent (PGD) algorithm. Starting from an initial estimate $\xv_0$, the iterations are given by
\begin{align}
    \sv_{t+1} &= \xv_t - \mu \nabla f_L(\xv_t), \label{eq:pgd_grad}\\
    \xv_{t+1} &= \Pi_{\Xc}(\sv_{t+1}), \label{eq:pgd_proj}
\end{align}
where $\mu>0$ denotes the step size, $\nabla f(\xv_t)$ is the gradient of the negative log-likelihood at iteration $t$, and $\Pi_{\Xc}(\cdot)$ denotes a projection onto the constraint set $\Xc$. The gradient descent step enforces data fidelity through the likelihood function, while the projection step incorporates prior information by restricting the iterate to the admissible set $\Xc$. 

In practice, $\Pi_{\Xc}(\cdot)$ can be implemented using a codebook or untrained neural networks such as implicit neural representations (INRs) \cite{sitzmann2020implicit} or deep image prior (DIP) \cite{ulyanov2018deep}, which parameterize images in the class $\Xc$. In this work, we adopt DIP as the projection operator. Specifically, DIP is represented by a neural network $g_{\theta}(\cdot)$, parameterized by $\theta$, that takes a fixed random input.

The projection is carried out by optimizing $\theta$ at each iteration $t$:
\begin{align*}
&\hat{\theta}_t = \operatorname*{argmin}_{\theta} \| g_{\theta}(\cdot) - \sv_{t+1} \|_2^2, \\
&\xv_{t+1} = g_{\hat{\theta}_t}(\cdot).
\end{align*}

Next, we describe how to efficiently compute $\nabla f_L(\xv_t)$, addressing the computational challenges associated with matrix inversion. The gradient of $f_L(\xv_t)$ in \eqref{eq:loss} with respect to $\xv_t$ at each iteration $t$ of PGD is given by (see \ref{sec:app_A} for derivation details)
\begin{align} \label{eq:grad}
    \nabla f_L(\xv_t) &
    = \diag (A^H S_t^{-1} A) - |A^H S_t^{-1} \yv_1|^2 + \sum_{\ell=2}^L \Bigg[ \diag (A^H M_t^{-1} A) - |A^H M_t^{-1} \rv_{t,\ell}|^2 \nonumber \\
    &- \alpha^2 \bigg( 2 \Re \Big\{ \diag (A^H S_t^{-1} B_t M_t^{-1} A) - A^H S_t^{-1} B_t M_t^{-1} \rv_{t,\ell} \odot \overline{A^H M_t^{-1} \rv_{t,\ell}} \Big\} \nonumber \\ 
    &- \diag (A^H S_t^{-1} B_t M_t^{-1} B_t S_t^{-1} A) + |A^HS_t^{-1} B_t M_t^{-1} \rv_{t,\ell}|^2 \bigg) \nonumber \\
    & + 2 \alpha \Re \Big\{ A^H B_t S_t^{-2} \yv_{\ell-1} \odot \overline{A^H M_t^{-1} \rv_{t,\ell}} - A^H S_t^{-1} \yv_{\ell-1} \odot \overline{A^H M_t^{-1} \rv_{t,\ell}} \Big\} \Bigg].
\end{align}
Note that $B(\xv_t), S(\xv_t), M(\xv_t), \rv_\ell(\xv_t)$ as functions of $\xv_t$ change through the PGD iteration, where we simplify the notation as $B_t, S_t, M_t, \rv_{t,\ell}$. We can see that the gradient vector $\nabla f(\xv_t)$ involves multiple matrix-vector and matrix-matrix multiplication, each of which requires one or more matrix inversion. The direct calculation of the gradient is particularly intensive, especially when signal dimension $n$ and the number of looks acquired $L$ are large, in the context of an iterative algorithm. In particular, repeated evaluations of $S_t^{-1}$ and $M_t^{-1}$ dominate the computational cost.

\subsection{Efficient Calculation of the Gradient}

To efficiently compute the gradient in \eqref{eq:grad}, we decompose the required terms into two categories: vector-valued terms and diagonal matrix terms. The vector terms include  $A^H S^{-1} \yv_1$, $A^H M^{-1} \rv_\ell$, $A^H S^{-1} B M^{-1} \rv_\ell$, $A^H B S^{-2} \yv_{\ell-1}$, $A^H S^{-1} \yv_{\ell-1}$, while the diagonal terms include  $\diag (A^H S^{-1} A)$, $\diag (A^H M^{-1} A)$, $\diag (A^H S^{-1} B M^{-1} A)$, $\diag (A^H S^{-1} B M^{-1} B S^{-1} A)$.

Direct computation of these quantities is computationally prohibitive, as it requires forming large matrices and computing their inverses. To address this, we first employ the conjugate gradient method in conjunction with the Fourier structure of $A$ to avoid explicit matrix formation and inversion. We then develop a Monte Carlo–based randomized approach to approximate the diagonal terms efficiently, where conjugate gradient is again used to avoid direct matrix inversion.

\subsubsection{Conjugate gradient for vector terms involving matrix inversion}

We first show how to compute $A^H S^{-1} \yv_1$ and $A^H M^{-1} \rv_\ell$, two representative vector terms, without explicitly forming $S^{-1}$ and $M^{-1}$. These computations provide the necessary building blocks for evaluating other similar vector terms.

To compute $A^H S_t^{-1} \yv_1$ in \eqref{eq:grad}, define $\hv_t = S_t^{-1} \yv_1$. Note that $\hv_t$ is the solution to the following least-squares problem:
\begin{align}\label{eq:cg_1}
    \min_{\hv} \| S_t \hv - \yv_1 \|^2.
\end{align}
Therefore, if we can efficiently solve \eqref{eq:cg_1}, we obtain the desired quantity $A^H S_t^{-1} \yv_1$ by applying $A^H$ to $\hv_t$.

To solve \eqref{eq:cg_1} efficiently, we employ the conjugate gradient method \cite{golub2013matrix}, which leverages the fact that $S_t$ is Hermitian positive definite. The details are provided in Algorithm~\ref{alg:CG}.

\begin{algorithm}[t]
\caption{Conjugate gradient for 
$\operatorname*{argmin}_{\mathbf{h}} 
\| S_t \mathbf{h} - \yv_1 \|^2$}
\label{alg:CG}
\begin{algorithmic}[1]

\STATE {\bfseries Initialize} 
$\hv_0=\mathbf{0}$, $\wv_0 = \yv_1 - S_t \hv_0$,
initial direction $\mathbf{p}_0 = \mathbf{r}_0$,
stopping tolerance $\epsilon$.

\FOR{$j = 0,1,\ldots,J-1$}
    \STATE $
    \alpha_j =
    \frac{\wv_j^{H}\wv_j}
         {\mathbf{p}_j^{H}S_t\mathbf{p}_j}
    $
    \hfill (step size)
    \STATE $
    \mathbf{h}_{j+1} = \mathbf{h}_j + \alpha_j \mathbf{p}_j
    $
    \hfill (solution update)
    \STATE $
    \wv_{j+1} = \wv_j - \alpha_j S_t\mathbf{p}_j
    $
    \hfill (residual update)
    \IF{$\|\wv_{j+1}\|_2 \le \epsilon$}
        \STATE $\hat{\mathbf{h}} = \mathbf{h}_{j+1}$ \textbf{break}
        \hfill (convergence)
    \ENDIF
    \STATE $
    \beta_j =
    \frac{\wv_{j+1}^{H}\wv_{j+1}}
         {\wv_j^{H}\wv_j}
    $
    \hfill (direction weight)
    \STATE $
    \mathbf{p}_{j+1} = \wv_{j+1} + \beta_j \mathbf{p}_j
    $
    \hfill (new search direction)
\ENDFOR
\STATE {\bfseries Output} $\hat{\mathbf{h}} = \mathbf{h}_J$
\end{algorithmic}
\end{algorithm}

Running Algorithm~\ref{alg:CG} requires computing expressions of the form $S_t \mathbf{p}_j$ and $S_t \hv_0$. Owing to the special structure of the forward operator, these operations can be performed efficiently using the Fourier transform. Specifically, for a 2D image $\Xb$ with vectorized form $\xv = \operatorname{vec}(\Xb)$,
\[
A \xv = \operatorname{vec}(\Ac(\Xb)), 
\]
where $\Ac(\Xb) = \operatorname{IFFT2} \;(P \odot \operatorname{FFT2}(\Xb))$.
Here, $\operatorname{FFT2}$ denotes the two-dimensional Fourier transform, and $P$ represents the 2D aperture mask defined earlier. Similarly, the matrix-vector product $S_t \hv$ can be computed by applying the operator $S(\xv_t)$ to the 2D representation of $\hv$, denoted by $\mathbf{H}$, as
\[
S(\xv_t)\hv
= \operatorname{vec} \left(\Ac \left(\Xb_t \odot \Ac(\mathbf{H})\right) + \sigma_z^2 \mathbf{H}\right).
\]
Here, $\Xb_t$ denotes the 2D image corresponding to the estimate $\xv_t$ at iteration $t$. By operating directly in the 2D domain, these computations can be performed efficiently without explicit formation of large matrices. Furthermore, by exploiting the structure of the Fourier optics operator in \eqref{eq:sensing_matrix} and leveraging fast Fourier transform (FFT) implementations within the conjugate gradient method, the overall reconstruction is significantly accelerated and scales efficiently to high-resolution images.

Next, to compute $A^H M_t^{-1} \rv_\ell$, first we need to obtain $\rv_{t,\ell} = \yv_\ell - \alpha B_t S_t^{-1} \yv_{\ell-1}$, in which the $S_t^{-1}$ is not needed as in \eqref{eq:cg_1}. Assuming we have access to $M_t$, we can avoid $M_t^{-1}$ by using conjugate gradient to solve the following least-squares
\begin{align}\label{eq:cg_2}
    \operatorname*{argmin}_{\gv} \| M_t \gv - \rv_\ell \|^2,
\end{align} 
where the solution $\hat{\gv}_t = M_t^{-1} \rv_\ell$. However, even $M_t$ requires the inversion $S_t^{-1}$, which we avoid computing and saving explicitly. To achieve this, we apply a nested CG to avoid computing $S_t^{-1}$ when applying $M_t$ in CG.

Given an arbitrary vector $\sv$, when it is multiplied with $M_t = S_t - \alpha^2 B_t S_t^{-1} B_t$ in conjugate gradient, we solve the least-squares to avoid $S_t^{-1}$
\begin{align}\label{eq:cg_3}
    \operatorname*{argmin}_{\qv} \| S_t \qv - B \sv \|^2.
\end{align} 
where the solution $\hat{\qv}_t = S_t^{-1} B \sv$ can be obtained with the nested CG. Having $\hat{\qv}_t$, the product $M_t \sv$ is expressed as $S_t \sv - \alpha^2 B_t \hat{\qv}_t$. The remaining direct vector terms in \eqref{eq:grad} are computed following the same methods used for $A^H S_t^{-1} \yv_1$ and $A^H M_t^{-1} \rv_{t,\ell}$.

\begin{algorithm}[t]
\caption{Monte-Carlo matrix diagonal approximation}
\label{alg:MC}
\begin{algorithmic}[1]
\STATE Given $A,S_t$, approximate $\diag (A^HS_t^{-1}A)$
\FOR{$k=1,2,\ldots,K$}
    \STATE $\vv^{(k)}_t \sim \Nc(0,I_n)$
    \STATE $\hat{\uv}^{(k)}_t \leftarrow 
    \operatorname*{argmin}_{\mathbf{c}} 
    \| S_t \mathbf{c} - A\vv^{(k)}_t \|^2$
    \hfill (Apply CG)
\ENDFOR
\STATE $\widehat{\diag (A^HS_t^{-1}A)} \leftarrow \frac{1}{K}\sum_{k=1}^K A^H\hat{\mathbf{c}}^{(k)}_t \odot \vv^{(k)}_t$

\end{algorithmic}
\end{algorithm}

\subsubsection{Monte-Carlo approximation of matrix diagonal}

Next, we employ a method based on randomized linear algebra to estimate the diagonal entries. In particular, we proposed an efficient approach for estimating $\diag (A^H S_t^{-1} A)$, and the other matrix diagonals in \eqref{eq:grad}. To this end, let $\vv \sim \Nc(0, I_n)$ and define $\hat{\mathbf{c}}_t$ by $\hat{\mathbf{c}}_t = S_t^{-1} A \vv$.
As before, $\hat{\mathbf{c}}_t$ can be computed as the solution to the following least-squares as in Algorithm \ref{alg:CG}
\begin{align} \label{eq:cg_2}
    \hat{\mathbf{c}}_t = \operatorname*{argmin}_{\mathbf{c}} \bigl\| S_t\mathbf{c} - A\vv \bigr\|^2 .
\end{align}
Note that, since the entries of $\vv$ are zero-mean and independent,
\begin{align}
    \E \left[A^H \hat{\mathbf{c}}_t \odot \vv\right]
    = \E \left[(A^H S_t^{-1} A \vv)\odot \vv\right]
\end{align}
is equal to the diagonal entries of $A^H S^{-1}_t A$. This observation implies that $A^H \hat{\mathbf{c}}_t \odot \vv$ provides an unbiased estimator of the desired diagonal entries. 

To exploit this property and efficiently estimate the diagonal entries, we adopt a Monte Carlo approach. Specifically, we independently generate $\vv^{(1)},\ldots,\vv^{(K)}$, each distributed as $\mathcal{N}(\mathbf{0}, I_n)$, and compute $\hat{\mathbf{c}}^{(k)}_t$, for $k=1,\ldots,K$, by solving the corresponding least-squares problems. The desired diagonal entries are then estimated as
\begin{align} \label{eq:MC}
    \frac{1}{K} \sum_{k=1}^K A^H \hat{\mathbf{c}}^{(k)}_{t} \odot \vv^{(k)}_{t}.
\end{align}
In practice, for sufficiently large $K$, this approximation yields an accurate estimate, as demonstrated in \cite{chen2026monte}. The algorithm is summarized as in Algorithm~\ref{alg:MC}. Overall, the computation follows a hierarchical structure: outer Monte Carlo loops generate probe vectors, inner CG solvers evaluate inverse operator actions, and operator compositions (e.g., $A$, $B$, $S$, $M$) define the transformation pipeline. 

This framework enables direct computation of the likelihood gradient without forming large matrices and performing expensive matrix–vector multiplications. It also avoids the memory overhead associated with storing such matrices. The resulting gradient descent step is efficient and accurate while maintaining scalability to high-dimensional imaging systems.

\begin{figure}[t]
    \centering

    {\footnotesize
    \subfloat[Circular aperture]{
        \includegraphics[width=0.3\columnwidth]{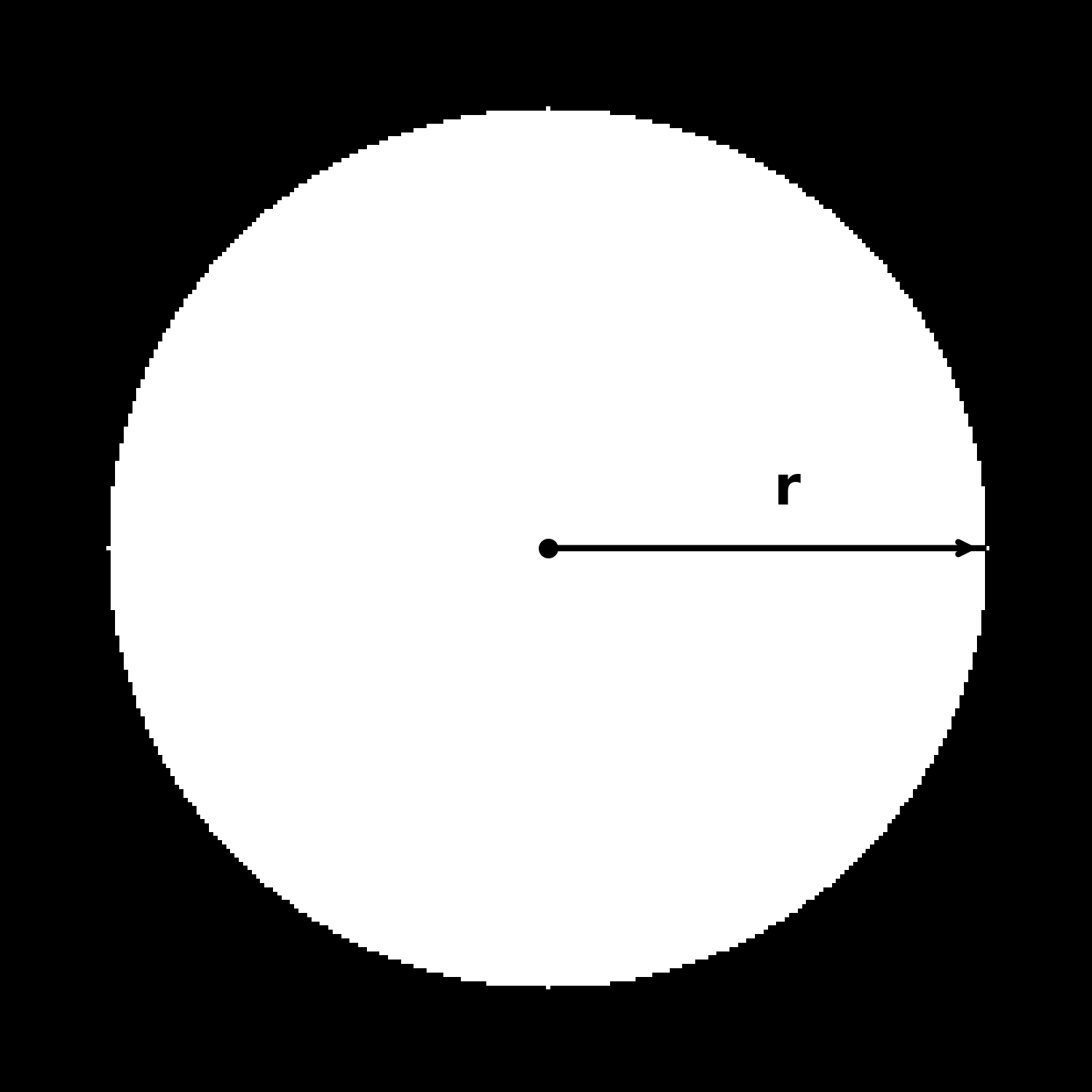}
    }
    \hspace{-0.4em}
    \subfloat[Annular aperture]{
        \includegraphics[width=0.3\columnwidth]{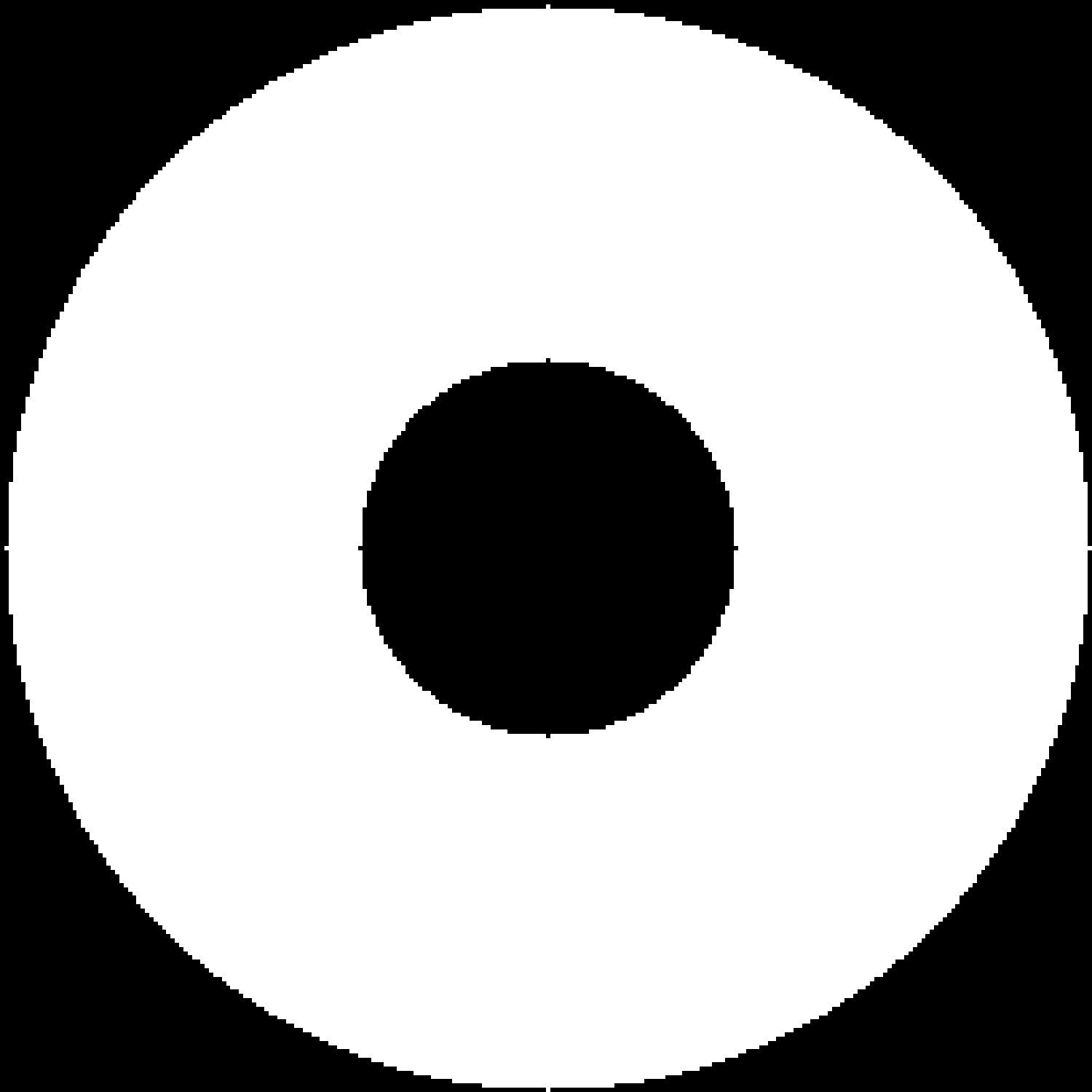}
    }
    }
    \caption{Apertures in the imaging forward model~\eqref{eq:forward_multi}.}
    \label{fig:apertures}

\end{figure}

\subsection{Estimation of Correlation Coefficient $\alpha$}

The proposed MLE-based optimization requires knowledge of the correlation coefficient $\alpha$. However, in practice, $\alpha$ is unknown and needs to be estimated based on the measurements.  From the measurement model \eqref{eq:forward_model_correlation}, it  follows that
\begin{align}
    \mathbb{E}[\yv_\ell \yv^H_\ell] &= AXA^H+\sigma^2_zI, \nonumber \\
    \mathbb{E}[\yv_{\ell-1} \yv^H_\ell] &= \alpha AXA^H. 
\end{align}
Therefore, 
\begin{align}
    \mathbb{E}[\|\yv_\ell\|_2^2] &= \tr(AXA^H)+ n\sigma^2_z, \nonumber \\
    \mathbb{E}[\yv_{\ell-1}^H \yv_\ell] &= \alpha \tr(AXA^H).
\end{align}
Let $\gamma=\tr(AXA^H)$. This observation suggests that, given  $\yv_1,\ldots,\yv_L$, we can estimate $\gamma$ and $\alpha$ as follows:
\begin{align}
    \hat{\gamma} = \frac{1}{nL}\sum^L_{\ell=1} \sum^n_{i=1} y_{\ell, i} \overline{y_{\ell, i}}.
\end{align}
In this approximation, we ignore the contribution of $\sigma_z^2$. As shown later in our simulation results, this simplification has a limited impact on estimation accuracy. Since
\[
\alpha = \frac{\mathbb{E}[\yv_{\ell-1}^H \yv_\ell]}{\tr(AXA^H)},
\]
it can estimated with $\hat{\gamma}$ as
\begin{align}
    \hat{\alpha} =\frac{1}{\hat{\gamma}}\left(\frac{1}{ n(L-1)}\sum^L_{\ell=2} \sum^n_{i=1} y_{\ell-1, i} \overline{y_{\ell,i}}\right).
\end{align}
The estimation accuracy is evaluated in Section~\ref{sec:4-1}.

\section{Simulations}\label{sec:sim}

In this section, we evaluate the performance of the proposed algorithm. In our simulations, correlated multi-look measurements are generated according to \eqref{eq:forward_model_correlation}. Our goal is to assess the impact of speckle correlation and to compare the performance of the proposed loss function with that of a baseline assuming fully independent measurements.

We have the following parameters in the experimental setup for exploration: correlation coefficient $\alpha$, number of looks $L$, additive noise level $\sigma_z$. 

When creating the aperture, $P$ corresponds to a circular aperture that acts as a low-pass filter in the spatial frequency domain and is defined as
\[
P(h,w) =
\begin{cases}
1, & \sqrt{(h-h_0)^2 + (w-w_0)^2} \le r,\\[4pt]
0, & \text{otherwise},
\end{cases}
\]
where $(h_0, w_0)$ denotes the aperture center and $r$ is the radius. To account for central obscuration encountered in practical optical systems, we also consider annular apertures \cite{goodman2005introduction}. We evaluate our algorithm on circular aperture with different radii $2r / H$, as well as the annular aperture with transparency ratio around $0.7$ shown in Figure~\ref{fig:apertures}.

We evaluate the algorithm from two perspectives, 1) reconstruction performance, measured in terms of  PSNR and SSIM; and 2) computational efficiency of the proposed gradient evaluation method. Prior to  these experiment, we first evaluate  accuracy of the proposed method for estimating  $\alpha$, which is necessary for the reconstruction algorithm.

\begin{table}[t] 
\small
% \tiny
% \scriptsize
\caption{The estimated correlation coefficient $\widehat{\alpha}$ over 50 runs: $\mathrm{mean} \pm \mathrm{std}$, under different system settings when the number of looks is $L=4$, image resolution is $256 \times 256$.} 
\begin{center}
% \begin{sc}
% \resizebox{\columnwidth}{!}{
\begin{tabular}{ccccc}
    \toprule
    Aperture $\frac{2r}{H}$ & $\sigma_z$ 
    & $\alpha=0.2$ & $\alpha=0.5$ & $\alpha=0.8$ \\
    \midrule
    \multirow{2}{*}{0.8}
        & 15 & $0.1967 \pm 0.0022$ & $0.4924 \pm 0.0019$  & $0.7883 \pm 0.0013$  \\
       & 25  & $0.1917 \pm 0.0022$ & $0.4800 \pm 0.0019$ & $0.7685 \pm 0.0014$  \\
    \cmidrule(l){1-5}

    \multirow{2}{*}{1.0}
       & 15  & $0.1982 \pm 0.0017$ & $0.4953 \pm 0.0014$ & $0.7926 \pm 0.0009$  \\
       & 25  & $0.1950 \pm 0.0017$ & $0.4873 \pm 0.0015$ & $0.7797 \pm 0.0010$  \\
    \bottomrule
\end{tabular}
% }
% \end{sc}
\label{tab:alpha_estimate}
\end{center}

\end{table}

\subsection{Correlation Coefficient Estimation}\label{sec:4-1}
We first empirically evaluate the accuracy of the correlation coefficient estimation under varying aperture radii $r$, noise levels $\sigma_z$, and correlation coefficients $\alpha$. As shown in Table~\ref{tab:alpha_estimate}, the estimation is accurate when the aperture transparency ratio is high and the noise level is low. Even under more challenging conditions, with lower transparency ratios and higher noise levels, the estimated values remain close to the true $\alpha$. Based on these observations, we adopt this estimation approach for use in the proposed reconstruction algorithm.

\begin{table*}[t] 
% \small
% \tiny
% \scriptsize
\caption{Reconstruction performance in terms of PSNR (dB) and SSIM under varying speckle correlation coefficients, aperture settings, numbers of looks, and noise levels. We compare the proposed loss function~\eqref{eq:loss} (using estimated $\hat{\alpha}$) with baseline \cite{chen2026monte} that use loss function~\eqref{eq:objective_independent} assuming independence ($\hat{\alpha}=0$). Results are reported for the \textit{Peppers} image.}
\begin{center}
% \begin{sc}
\resizebox{\textwidth}{!}{
\begin{tabular}{ccccccccccccc}
    \toprule
    \multicolumn{1}{c}{Aperture} & 
    \multicolumn{1}{c}{Noise level} & 
    \multicolumn{1}{c}{Lower bound} & 
    \multicolumn{1}{c}{Looks} &
    \multicolumn{1}{c}{$\alpha = 0.0$} &
    \multicolumn{2}{c}{$\alpha = 0.2$} & 
    \multicolumn{2}{c}{$\alpha = 0.5$} &
    \multicolumn{2}{c}{$\alpha = 0.8$} 
    \\
    \cmidrule(lr){1-1}
    \cmidrule(lr){2-2}
    \cmidrule(lr){3-3}
    \cmidrule(lr){4-4}
    \cmidrule(lr){5-5}
    \cmidrule(lr){6-7}
    \cmidrule(lr){8-9}
    \cmidrule(lr){10-11}

    $M$ & $\sigma_z$ & $L=1$ & $L$ 
    & Upper bound 
    &  \cite{chen2026monte} w/ \eqref{eq:objective_independent}
    & Ours w/ \eqref{eq:loss} 
    &  \cite{chen2026monte} w/ \eqref{eq:objective_independent}
    & Ours w/ \eqref{eq:loss} 
    &  \cite{chen2026monte} w/ \eqref{eq:objective_independent}
    & Ours w/ \eqref{eq:loss} \\
    \midrule

\multirow{6}{*}{\shortstack{Circular \\ $\frac{2r}{H}=0.8$}}
    & \multirow{3}{*}{15}
    & \multirow{3}{*}{19.21 / 0.4813} 
    & 2 & 20.45 / 0.5415 & 20.42 / 0.5413 
    & 20.43 / 0.5403 & 20.07 / 0.5178 
    & 20.43 / 0.5437 & 19.47 / 0.4805 
    & 20.11 / 0.5364 \\
    &   &  
    & 4 & 21.48 / 0.5888 & 21.39 / 0.5842 
    & 21.45 / 0.5882 & 20.94 / 0.5553 
    & 21.36 / 0.5861 & 19.76 / 0.4776 
    & 20.84 / 0.5798 \\
    &   &  
    & 10 & 22.65 / 0.6615 & 22.55 / 0.6545 
    & 22.57 / 0.6609 & 22.12 / 0.6217 
    & 22.42 / 0.6577 & 20.81 / 0.5242 
    & 21.51 / 0.6395 \\
    \cmidrule(l){2-11}

    & \multirow{3}{*}{25}
    & \multirow{3}{*}{19.17 / 0.4780} 
    & 2 & 20.53 / 0.5430 & 20.45 / 0.5366 
    & 20.50 / 0.5407 & 20.18 / 0.5209 
    & 20.44 / 0.5411 & 19.57 / 0.4821 
    & 19.94 / 0.5301 \\
    &   &  
    & 4 & 21.54 / 0.5894 & 21.47 / 0.5838 
    & 21.54 / 0.5878 & 21.00 / 0.5528 
    & 21.39 / 0.5879 & 19.85 / 0.4813 
    & 20.46 / 0.5756 \\
    &   &  
    & 10 & 22.70 / 0.6576 & 22.67 / 0.6558 
    & 22.70 / 0.6593 & 22.22 / 0.6177 
    & 22.41 / 0.6573 & 20.94 / 0.5292 
    & 20.92 / 0.6372 \\

    \midrule

\multirow{6}{*}{\shortstack{Circular \\ $\frac{2r}{H}=1.0$}} 
    & \multirow{3}{*}{15}
    & \multirow{3}{*}{19.82 / 0.5004} 
    & 2 & 21.29 / 0.5642 & 21.27 / 0.5639 
    & 21.31 / 0.5655 & 20.96 / 0.5470 
    & 21.27 / 0.5674 & 20.34 / 0.5137 
    & 21.18 / 0.5658 \\
    &   &  
    & 4 & 22.34 / 0.6153 & 22.22 / 0.6101 
    & 22.29 / 0.6143 & 21.66 / 0.5736 
    & 22.24 / 0.6096 & 20.52 / 0.5077 
    & 21.69 / 0.5988 \\
    &   &  
    & 10 & 23.31 / 0.6804 & 23.26 / 0.6774 
    & 23.29 / 0.6805 & 22.81 / 0.6417 
    & 23.30 / 0.6814 & 21.64 / 0.5625 
    & 22.79 / 0.6633 \\
    \cmidrule(l){2-11}

    & \multirow{3}{*}{25}
    & \multirow{3}{*}{19.81 / 0.4993} 
    & 2 & 21.27 / 0.5630 & 21.27 / 0.5606 
    & 21.29 / 0.5641 & 20.99 / 0.5470 
    & 21.35 / 0.5661 & 20.42 / 0.5155 
    & 21.17 / 0.5628 \\
    &   &  
    & 4 & 22.34 / 0.6141 & 22.26 / 0.6095 
    & 22.33 / 0.6119 & 21.79 / 0.5769 
    & 22.29 / 0.6106 & 20.73 / 0.5201 
    & 21.95 / 0.6009 \\
    &   &  
    & 10 & 23.42 / 0.6852 & 23.24 / 0.6744 
    & 23.35 / 0.6782 & 22.94 / 0.6456 
    & 23.39 / 0.6809 & 21.71 / 0.5625 
    & 22.80 / 0.6603 \\

    \midrule

\multirow{6}{*}{\shortstack{Annular}}
    & \multirow{3}{*}{15}
    & \multirow{3}{*}{19.62 / 0.4937} 
    & 2 & 21.05 / 0.5631 & 21.02 / 0.5619 
    & 21.06 / 0.5632 & 20.69 / 0.5422 
    & 21.06 / 0.5671 & 20.10 / 0.5083 
    & 20.84 / 0.5577 \\
    &   &  
    & 4 & 22.13 / 0.6121 & 22.00 / 0.6052 
    & 22.13 / 0.6118 & 21.61 / 0.5789 
    & 22.00 / 0.6088 & 20.47 / 0.0000 
    & 21.72 / 0.6053 \\
    &   &  
    & 10 & 23.21 / 0.6777 & 23.17 / 0.6701 
    & 23.26 / 0.6816 & 22.79 / 0.6398 
    & 23.15 / 0.6780 & 21.51 / 0.5557 
    & 22.56 / 0.6616 \\
    \cmidrule(l){2-11}

    & \multirow{3}{*}{25}
    & \multirow{3}{*}{19.61 / 0.4923} 
    & 2 & 21.12 / 0.5640 & 21.04 / 0.5583 
    & 21.10 / 0.5619 & 20.76 / 0.5425 
    & 21.08 / 0.5621 & 20.17 / 0.5061 
    & 20.82 / 0.5578 \\
    &   &  
    & 4 & 22.19 / 0.6138 & 22.08 / 0.6070 
    & 22.19 / 0.6136 & 21.64 / 0.5764 
    & 22.09 / 0.6087 & 20.52 / 0.5122 
    & 21.56 / 0.5996 \\
    &   &  
    & 10 & 23.31 / 0.6801 & 23.21 / 0.6723 
    & 23.27 / 0.6774 & 22.84 / 0.6418 
    & 23.19 / 0.6781 & 21.56 / 0.5556 
    & 22.33 / 0.6589 \\

    \bottomrule
\end{tabular}
}

% \end{sc}
\label{tab:main_results}
\end{center}
\end{table*}

\subsection{Reconstruction Performance}

The main objective of our experiments is to compare the reconstruction performance achieved by two loss functions: the proposed formulation in \eqref{eq:loss}, which accounts for inter-look correlation, and a baseline that assumes independence across looks in \cite{chen2026monte}.

In our model, larger values of $\alpha$ correspond to stronger speckle correlation. In the extreme case of $\alpha = 0$, the looks are fully independent, which serves as an upper performance bound. In contrast, when $\alpha = 1$, all looks are identical, effectively reducing the problem to a single-look setting ($L = 1$); this represents a lower performance bound under the same aperture radius ratio $\frac{2r}{H}$ and noise level $\sigma_z$.

The main results are reported in Table~\eqref{tab:main_results}. As $\alpha$ increases, across all aperture settings and noise levels, the proposed loss function consistently outperforms the baseline that assumes full independence. Even in the high-correlation regime (e.g., $\alpha = 0.8$), the PSNR achieved by the proposed method remains within $0.5$ dB of the performance upper bound.

\begin{figure*}[t]
    \centering
    
    \begin{subfigure}{0.825\linewidth}
        \centering
        \includegraphics[width=1.0\linewidth]{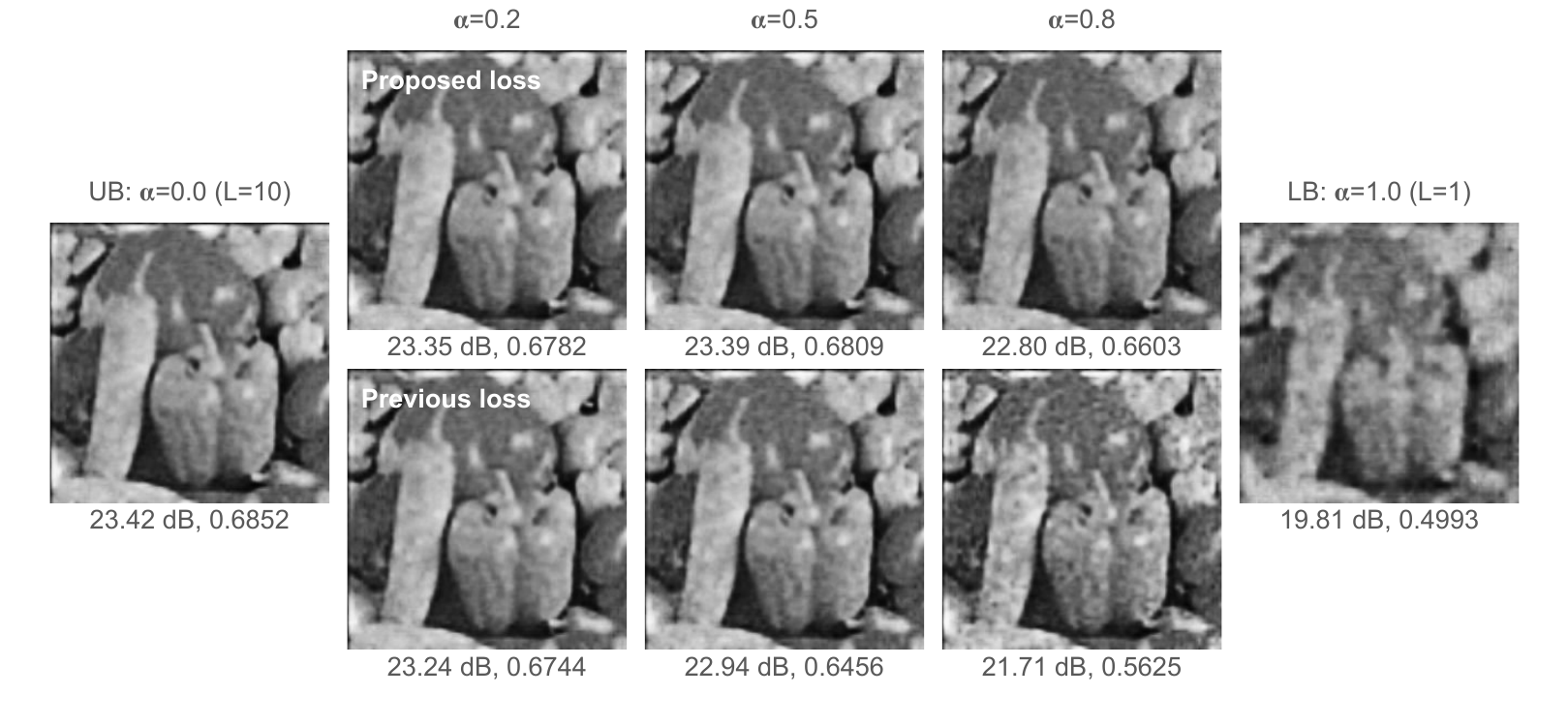}
        \caption{Reconstructed test image \textit{Peppers}.}
    \end{subfigure}

    \begin{subfigure}{0.825\linewidth}
        \centering
        \includegraphics[width=1.0\linewidth]{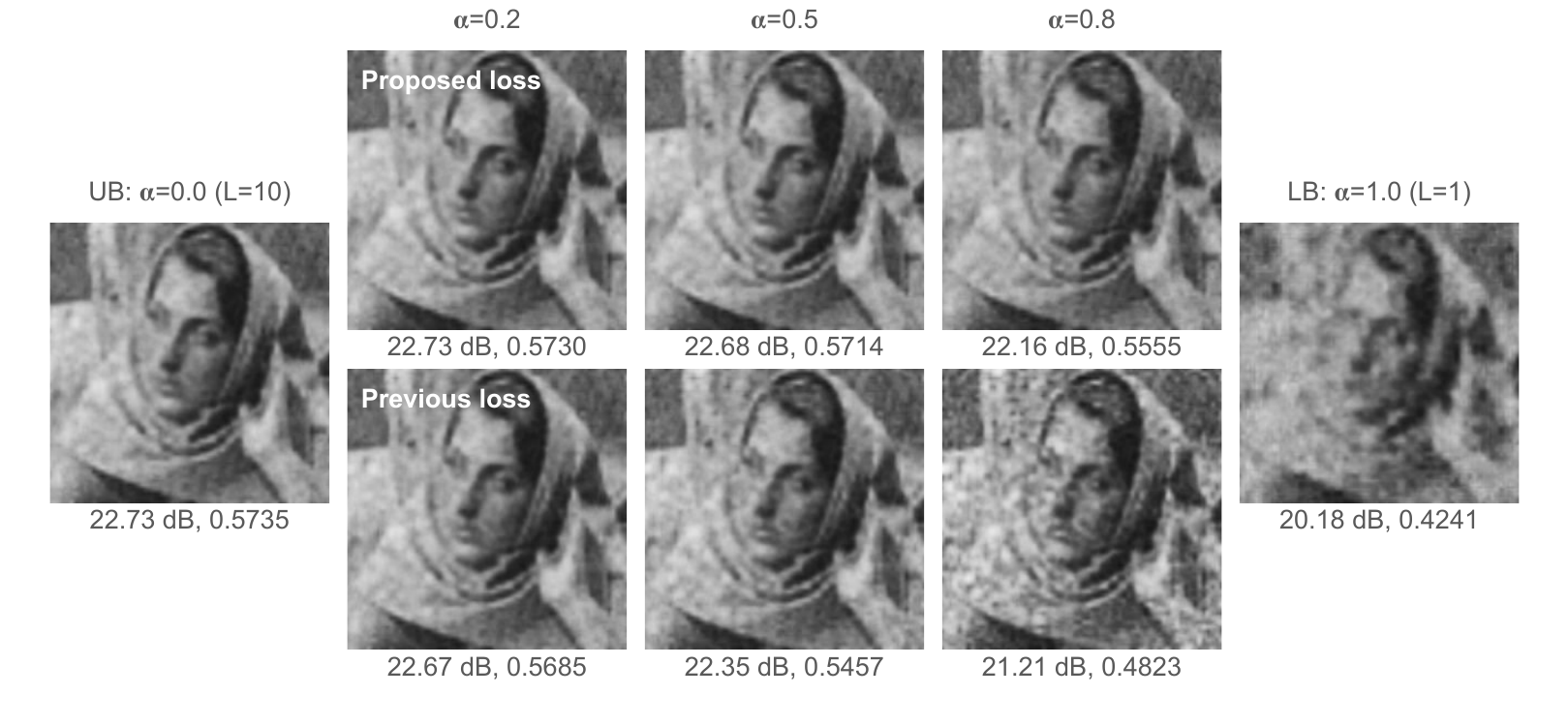}
        \caption{Reconstructed test image \textit{Barbara}.}
    \end{subfigure}
    
    \caption{The images reconstructed based on two different loss functions \eqref{eq:objective_independent} and \eqref{eq:loss}, when the correlation coefficient $\alpha$ takes the values $0.2, 0.5, 0.8$.}
    \label{fig:vis_alpha}
\end{figure*}

Figure~\ref{fig:vis_alpha} presents visual comparisons of reconstructed images obtained using the proposed loss function and the baseline assuming independence, under different values of $\alpha$ in the measurement model.

\begin{figure}[t]
\centering
\includegraphics[width=0.48\textwidth]{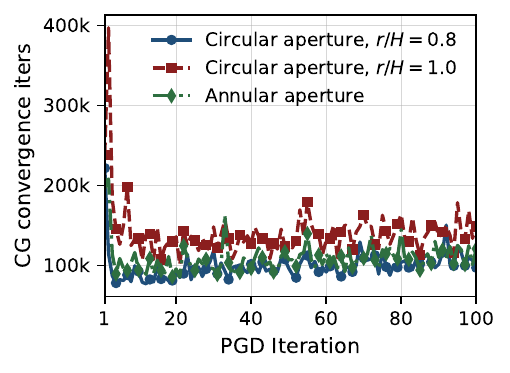}
\caption{Total CG iterations required at each PGD iteration for three different apertures. Monte-Carlo samples $K=50$, number of looks $L=10$.}
\label{fig:CG_iter}
\end{figure}

\subsection{Reconstruction Efficiency of the Algorithm}

We now evaluate the computational efficiency of the PGD. The gradient computation involves solving least-squares problems using the conjugate gradient (CG), where the number of CG convergence iterations directly impacts the runtime of each PGD iteration. 

Neglecting lower-order operations (e.g., inner products $\wv_j^{H}\wv_j$ and vector updates), the dominant computational cost arises from repeated applications of the operator $S$ (e.g., $S_t \mathbf{p}_j$) within the CG iterations. Since both direct and nested linear solves ($M_t$) rely on CG procedures involving $S$, the overall computational complexity is effectively characterized by the total number of $S$-operator applications.

In Figure~\ref{fig:CG_iter}, we report the number of CG iterations required to approximate $S^{-1}$ at each PGD iteration, which corresponds to the number of times the $S$-operator is applied. The number of the Monte-Carlo samples is $K=50$, the number of looks is $L=10$. We consider different apertures in the forward imaging process, including circular apertures with varying transparency ratios and annular apertures.

The $S$-operator consists of Fourier transform operations, which are efficiently implemented using FFT. For a $256 \times 256$ image, a single application of the $S$-operator takes approximately $2 \times 10^{-4}$ seconds. So, every iteration of the PGD algorithm takes approximately $20$ to $40$ seconds. The PGD algorithm takes $50$ to $100$ iterations to converge.

\section{Conclusion}

In this work, we investigate the reconstruction of speckle-free reflectivity from complex-valued multi-look measurements in digital holography, where the common assumption of independent speckle realizations across looks may be violated due to hardware constraints. We model the inter-look dependence using a first-order Markov process and derive the corresponding negative log-likelihood, leading to a constrained maximum likelihood estimation problem. To solve this problem efficiently, we adopt a PGD framework with implicit regularization via DIP. The computational challenges in evaluating the gradient are addressed using Monte Carlo–based diagonal approximation and conjugate gradient methods to avoid explicit matrix inversion, together with matrix-free implementations enabled by the Fourier optics structure. Simulation results demonstrate that the proposed approach remains robust under strong inter-look correlation, achieving performance close to the ideal independent-look scenario and outperforming methods that ignore such dependencies.

As future work, we plan to apply the proposed framework to real digital holography data, where the measurement process may deviate from the assumed model. In particular, real-world data may exhibit model mismatches, such as imperfect calibration, non-Gaussian noise, or more complex correlation structures. Extending the proposed method to handle such discrepancies and validating its performance on experimental datasets will be an important direction for further study.

\section*{Acknowledgment}
\noindent X.C., A.M., S.J. were supported in part by ONR award no. N00014-23-1-2371. S.J. was supported in part by NSF CCF-2237538.

\appendix
\renewcommand{\thesection}{Appendix \Alph{section}}
\section{Derivation of Gradient}
\label{sec:app_A}
We first compute the gradient with respect to $B$, and then use chain rule to obtain the gradient w.r.t $\xv$. By matrix identities and symmetry \cite{petersen2008matrix},
\begin{align}
d \log \det S &= \mathrm{tr}(S^{-1} dB), \\
d \big( \yv^H S^{-1} \yv \big)
&= - \mathrm{tr}( S^{-1} \yv \yv^H S^{-1} dB ).
\end{align}
Next we derive the gradient of $f_L$ with respect to $B$, first  for the term corresponding to  $\ell=1$, and then for the other terms that correspond to  $\ell\geq 2$.

For the first term, i.e., $\log \det S
+ \yv_1^H S^{-1} \yv_1$, we have
\begin{equation}
G^{(1)} 
=
\frac{\partial }{\partial B}(\log \det S
+ \yv_1^H S^{-1} \yv_1)
=
S^{-1} - S^{-1} \yv_1 \yv_1^H S^{-1}.
\end{equation}

For the  terms corresponding to $\ell>1$, fixing $\rv_\ell$,  the differential of $M$ can be written as
\begin{equation}
dM = dB -
\alpha^2
\Big(
dB\, S^{-1} B
+ B S^{-1}\, dB
- B S^{-1}\, dB\, S^{-1} B
\Big).
\end{equation}
Therefore, the contribution through $M$ from $\log \det M
+ \rv_\ell^H M^{-1} \rv_\ell$
\begin{align}
d(\log \det M
+ \rv_\ell^H M^{-1} \rv_\ell) 
&=\mathrm{tr}( W_\ell\, dM ) \\
&=
\mathrm{tr} \bigg(
\Big[
W_\ell
- \alpha^2 (S^{-1} B W_\ell + W_\ell B S^{-1}) + \alpha^2 S^{-1} B W_\ell B S^{-1}
\Big] dB \bigg).
\end{align}
where
\begin{align}
W_\ell \triangleq M^{-1} - M^{-1} \rv_\ell (M^{-1} \rv_\ell)^H
\end{align}

On the other hand, for a fixed  $M$, the differential of $\rv_\ell$ can be written as
\begin{equation}
d \rv_\ell
=
\alpha (B S^{-1} - I) dB\, S^{-1} \yv_{\ell-1}.
\end{equation}
Therefore, given that $B$ is Hermitian, the contribution through $\rv_\ell$ from the quadratic term $\rv_\ell^H M^{-1} \rv_\ell$ yields
\begin{align}
d(\rv_\ell^H M^{-1} \rv_\ell) 
&= 2 \Re \{ (M^{-1} \rv_\ell)^H d \rv_\ell \} \nonumber \\
&= \mathrm{tr}\!\left(
\alpha (C_\ell dB + (C_\ell dB)^H)
\right) \nonumber \\
&= \mathrm{tr} \left(
\alpha (C_\ell + C_\ell^H) dB
\right),
\end{align}
where
\begin{align}
C_\ell \triangleq (B S^{-1} - I)\,  S^{-1} \yv_{\ell-1} (M^{-1} \rv_\ell)^H.
\end{align}
Finally, we compute the gradient with respect to $\xv$. Note that
\begin{equation}
dB = A \, \mathrm{diag}(d\xv) \, A^H.
\end{equation}
Therefore, since
\begin{equation}
\nabla f_L(\xv)
=
\mathrm{diag}\!\left(
A^H
\frac{\partial f_L}{\partial B}
A
\right),
\qquad
\frac{\partial f_L}{\partial x_i}
=
a_i^H
\frac{\partial f_L}{\partial B}
a_i,
\end{equation}
using the chain rule, it follows that
\begin{align}
&\frac{\partial f_L}{\partial B}
=
S^{-1} - S^{-1} \yv_1 \yv_1^H S^{-1} \nonumber
\\
&+\sum_{\ell=2}^L
\bigg(
W_\ell
- \alpha^2 (S^{-1} B W_\ell + W_\ell B S^{-1}) + \alpha^2 S^{-1} B W_\ell B S^{-1}
+ \alpha (C_\ell + C_\ell^H),
\bigg).
\end{align}
 where $W_\ell$ and $C_\ell$ are defined as
\begin{align*}
%B &\triangleq A X A^H, \\
%S &\triangleq B + \sigma_z^2 I, \\
%M &\triangleq S - \alpha^2 B S^{-1} B, \\
%\rv_\ell &\triangleq \yv_\ell - \alpha B S^{-1} \yv_{\ell-1}, \\
W_\ell &\triangleq M^{-1} - M^{-1} \rv_\ell (M^{-1} \rv_\ell)^H, \\
C_\ell &\triangleq (B S^{-1} - I)\,  S^{-1} \yv_{\ell-1} (M^{-1} \rv_\ell)^H.
\end{align*}
Therefore, the gradient $\nabla f_L(\xv)$ can be calculated as

\begin{align*}
    \nabla f_L(\xv) &= \diag (A^H S^{-1} A) - |A^H S^{-1} \yv_1|^2 + \sum_{\ell=2}^L \Bigg[ \diag (A^H M^{-1} A) - |A^H M^{-1} \rv_\ell|^2 \\
    &- \alpha^2 \bigg( \diag (A^H S^{-1} B M^{-1} A) - \diag (A^H S^{-1} B M^{-1} \rv_\ell \rv_\ell^H M^{-1} A) + \diag (A^H M^{-1} B S^{-1} A) \\
    &- \diag (A^H M^{-1} \rv_\ell \rv_\ell^H M^{-1} B S^{-1} A) - \diag (A^H S^{-1} B M^{-1} B S^{-1} A) + |A^HS^{-1}B M^{-1} \rv_\ell|^2 \bigg) \\
    & + 2 \alpha \Re \Big\{ \diag(A^H B S^{-2} \yv_{\ell-1} \rv^H_\ell M^{-1} A) - \diag(A^H S^{-1} \yv_{\ell-1} \rv^H_\ell M^{-1} A) \Big\} \Bigg] \\
    &= \diag (A^H S^{-1} A) - |A^H S^{-1} \yv_1|^2 + \sum_{\ell=2}^L \Bigg[ \diag (A^H M^{-1} A) - |A^H M^{-1} \rv_\ell|^2 \\
    &- \alpha^2 \bigg( 2 \Re \Big\{ \diag (A^H S^{-1} B M^{-1} A) - A^H S^{-1} B M^{-1} \rv_\ell \odot \overline{A^H M^{-1} \rv_\ell} \Big\} \\ 
    &- \diag (A^H S^{-1} B M^{-1} B S^{-1} A) + |A^HS^{-1}B M^{-1} \rv_\ell|^2 \bigg) \\
    & + 2 \alpha \Re \Big\{ A^H B S^{-2} \yv_{\ell-1} \odot \overline{A^H M^{-1} \rv_\ell} - A^H S^{-1} \yv_{\ell-1} \odot \overline{A^H M^{-1} \rv_\ell} \Big\} \Bigg].
\end{align*}

\newpage
\bibliographystyle{IEEEtran}
\bibliography{references}

\end{document}